\documentclass[10pt,aps,pra,twocolumn,floatfix]{revtex4}
\usepackage{graphicx}

\begin{document}
\title{Current developments in silicene and germanene}
\author{T. P. Kaloni$^1$}
\email{thaneshwor.kaloni@umanitoba.ca}
\author{G. Schreckenbach$^1$}
\author{M. S. Freund$^1$}
\author{U. Schwingenschl\"ogl$^2$}
\email{udo.schwingenschlogl@kaust.edu.sa}
\affiliation{$^1$Department of Chemistry, University of Manitoba, Winnipeg, MB, R3T 2N2, Canada}
\affiliation{$^2$KAUST, PSE Division, Thuwal 23955-6900, Saudi Arabia}

\begin{abstract}
Exploration of the unusual properties of the two-dimensional materials silicene and
germanene is a very active research field in recent years. This article therefore
reviews the latest developments, focusing both on the fundamental materials properties
and on possible applications.
\end{abstract}

\maketitle
\section{Introduction}
Graphene, a single layer of C atoms closely packed in a two-dimensional (2D) honeycomb
lattice, exhibits intriguing electronic properties such as a very high carrier mobility
\cite{Novoselov}. The valence and conduction bands form a Dirac cone at the Fermi energy
(zero gap semiconductor) with the electrons behaving as massless Dirac fermions,
which can be useful in nanoelectronics \cite{Xu}, for example. However, difficulties
in the realization and tuning of a reasonable band gap currently are shifting interest
from graphene to other 2D materials, such as $h$-BN \cite{Watanabe,Kubota,Meyer},
transition metal dichalcogenides \cite{Novoselov3,Geim2,sant1,sant2}, silicene 
\cite{sahin1,Fleurence,datta,Meng,Xu,Quaresima,Chen}, germanene
\cite{Lia,kaloni-jap,NJP,nano-2015}, and silica \cite{Ongun}. 
Being composed of the group-IV elements Si and Ge, silicene and germanene immediately
come into mind as alternatives to graphene. Silicene has been predicted theoretically
by Takeda and Shiraishi in 1994 \cite{Takeda}, re-investigated by Guzm\'an-Verri
and co-workers in 2007 \cite{Voon}, and confirmed to be stable by Cahangirov and
co-workers in 2009 \cite{ciraci3}. Subsequently various groups have grown it on
different Ag surfaces \cite{Quaresima,Chen,n1} and other metallic substrates.

Density functional theory has been employed in order to understand the structural
and electronic properties of silicene on a series of metallic and semiconducting
substrates \cite{Lin2,Rubio,Wang,Quhe,kaloni-sci,Liu1}. It predicts for pristine
silicene a Dirac cone with $\pi$ and $\pi^*$ bands crossing at the K/K$'$-points
of the Brillouin zone \cite{Lebegue}. Stability has been demonstrated even
under high tensile strain of 17-20\% \cite{liu-epl,udo1,Dresselhaus1}. An 
external electric field can be used to open a tunable band gap \cite{Drummond,Ni} and
combination with the strong spin orbit coupling leads to novel phases that are 
interesting for spintronics devices \cite{prb-ezawa,iop,pccp}. A sizable
band gap can be achieved by hydrogenation \cite{Houssa-apl,jacob-prb,Hussain-pccp},
oxidation \cite{Ouyang-epl}, and chlorination \cite{jacob-prb}
as well as by molecule absorption \cite{freund}. Moreover, the electronic
structure shows strong modifications under Li, Na, K, Rb, and Cs decoration
\cite{Fei}. Heavy metals (including Au, Hg, Tl, and  Pb) can enhance the spin orbit
coupling \cite{kaloni-rrl} and new electronic phases have been found for transition
metal decoration \cite{Lin-prb,Zhao-prb,Zhang-sci}, such as the
quantum anomalous Hall state in the case of Co-decoration \cite{kaloni-prb14}.
Of particular interest are likewise the optical properties. For example, the infrared
absorbance shows a universal behaviour for all group-IV elements \cite{n8} and
many-body effects become important \cite{n9}. Both silicene and germanene reveal
distinct excitonic resonances, which are promising from the application point of
view, particularly after functionalization with H \cite{n10}.

Similar to silicene, pristine germanene is predicted to realize a corrugated structure
\cite{ciraci3} and to remain stable under high strain \cite{kaloni-cpl,Stein-prb,Cai-prb,Roome}.
Halogenated germanene has a substantial band gap \cite{ma1}, which can be further
widened by an external electric field \cite{Kusmartsev,Chen-jpc}, and it has been predicted that also a room temperature 2D topological phase can be achieved by functionalization \cite{si1}.
Absorption of alkali, alkali-earth, group-III, and 3$d$ transition metal atoms
can lead to metallic, half-metallic, and semiconducting states
\cite{Zhang-pccp,kaloni-jpcc}. Quantum spin Hall \cite{Dai-jmc,Seixas}, quantum
anomalous Hall \cite{Huang-prb,kaloni-jpcc}, and quantum spin/valley effects
\cite{Tabert} have been predicted theoretically. In addition, giant magnetoresistance
\cite{Ezawa-prb} and a large thermoelectric figure of merit (up to 2.5) at room
temperature \cite{Yang-prb} are expected. Germanene has been synthesized
on Al(111) and Au(111) substrates \cite{NJP,nano-2015}.
By interaction with Ag and $h$-BN substrates the Dirac cone will
disappear \cite{Zhao-pccp}, while the coupling to GaAs(0001) can be significantly
reduced by H intercalation \cite{kaloni-jap}. Graphene also has been put forward as
potential substrate \cite{Cai-prb}. Hydrogenated multilayer germanene experimentally
shows a band gap of 1.55 eV \cite{Bianco}.

\begin{table*}[ht]
\begin{tabular}{c||c|c|c|c|c}
& Lattice constant (\AA) & Bond length (\AA)& Buckling (\AA) & Band gap (meV)& Cohesive energy (eV/atom)\\ 
\hline\hline
graphene&2.45-2.47 \cite{neto,Wehling}&1.41-1.43 \cite{neto,Wehling} &0 \cite{xiao}   &0 \cite{xiao}  &7.37-7.81 \cite{Koskinen}\\
\hline
silicene&3.86-3.91 \cite{sahin1,peeters,datta}&2.22-2.26 \cite{sahin1,datta} &0.45-0.48 \cite{sahin1,peeters,ciraci3}  &1.55-2.00 \cite{kaloni-sci,Ni,Drummond}&4.69-5.06 \cite{Kwon}\\
\hline
germanene&4.06-4.10 \cite{Li,Scalise}&2.41-2.45 \cite{Li,Scalise} &0.66-0.68 \cite{ciraci3,kaloni-jap,Li}&24-25 \cite{Jiang,Seixas}&3.09 \cite{Koskinen}\\
\end{tabular}
\caption{Material parameters of graphene, silicene, and germanene.}
\end{table*}

Basic material parameters of graphene, silicene, and germanene
are compared in Table I. The buckling is defined as the maximal perpendicular distance
of atoms and increases from C to Si to Ge, while the cohesive energy decreases.
We note that MoS$_2$-like and dumbbell structures of 2D Si and Ge have been found to
be energetically favorable over graphene-like slightly buckled structures \cite{n2,n3}.
In the following we address significant theoretical and experimental developments
and insights in the physics and chemistry of silicene and germanene, reported mainly
in the last two years. 
 
\section{Silicene}
\subsection{Vibrational and electronic properties} 
In pristine silicene the spin orbit coupling results in a band gap of 2.0 meV, see
Fig.\ 1, which is larger than in graphene but smaller than in germanene and tinene
\cite{n11}. The band gap shrinks under biaxial tensile strain \cite{udo1}. In addition,
up to a strain of more than 5\% the Dirac cone is located at the Fermi energy, whereas
afterwards hole doping is introduced. At 10\% strain, for example, the 
Dirac point is located 0.18 eV above the Fermi energy. Comparison of the phonon
spectra of silicene without strain and under 10\% strain
in Fig.\ 2 shows that the frequencies of the G and D modes are strongly reduced.
However, only at large strain above 20\% the lattice becomes instable.
In Ref.\ \cite{PRB2013} the authors have demonstrated an anomalous thermal
response under uniaxial tensile strain, very different to graphene.
The thermal conductivity increases monotonically, which is
also true for nanoribbons, while the group velocities of the out-of-plane phonons
are suppressed.

\begin{figure}[t]
\includegraphics[width=0.48\textwidth,clip]{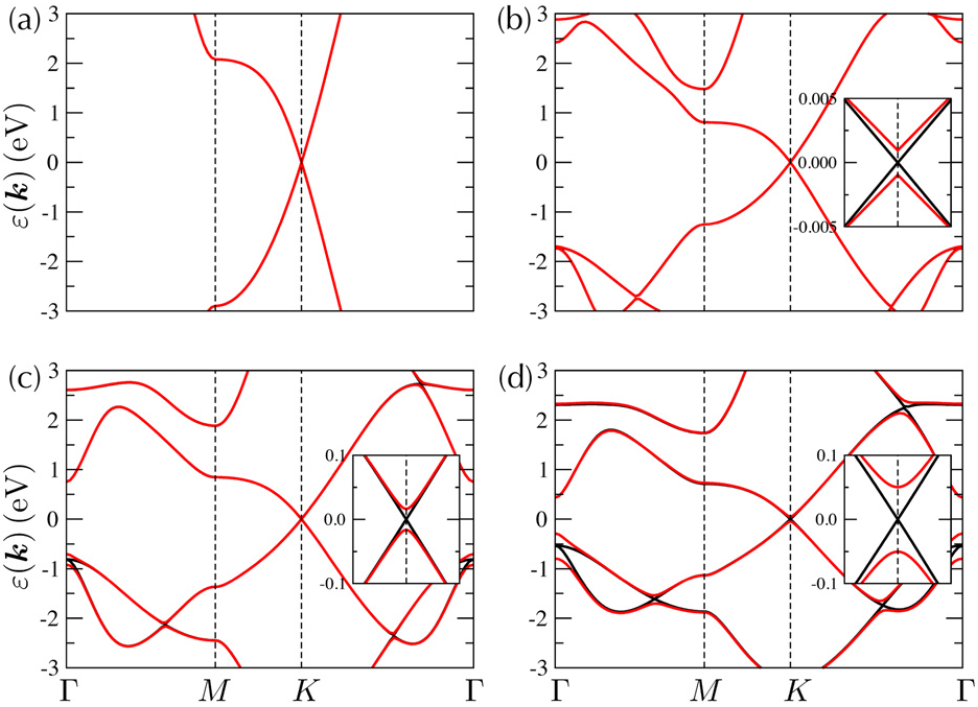}
\caption{Band structures of (a) graphene, (b) silicene, (c) germanene, and (d) tinene
obtained without (black lines) and with (red lines) spin orbit coupling (courtesy of
Ref.\ \cite{n11}).}
\end{figure}

\begin{figure}[t]
\includegraphics[width=0.35\textwidth,clip]{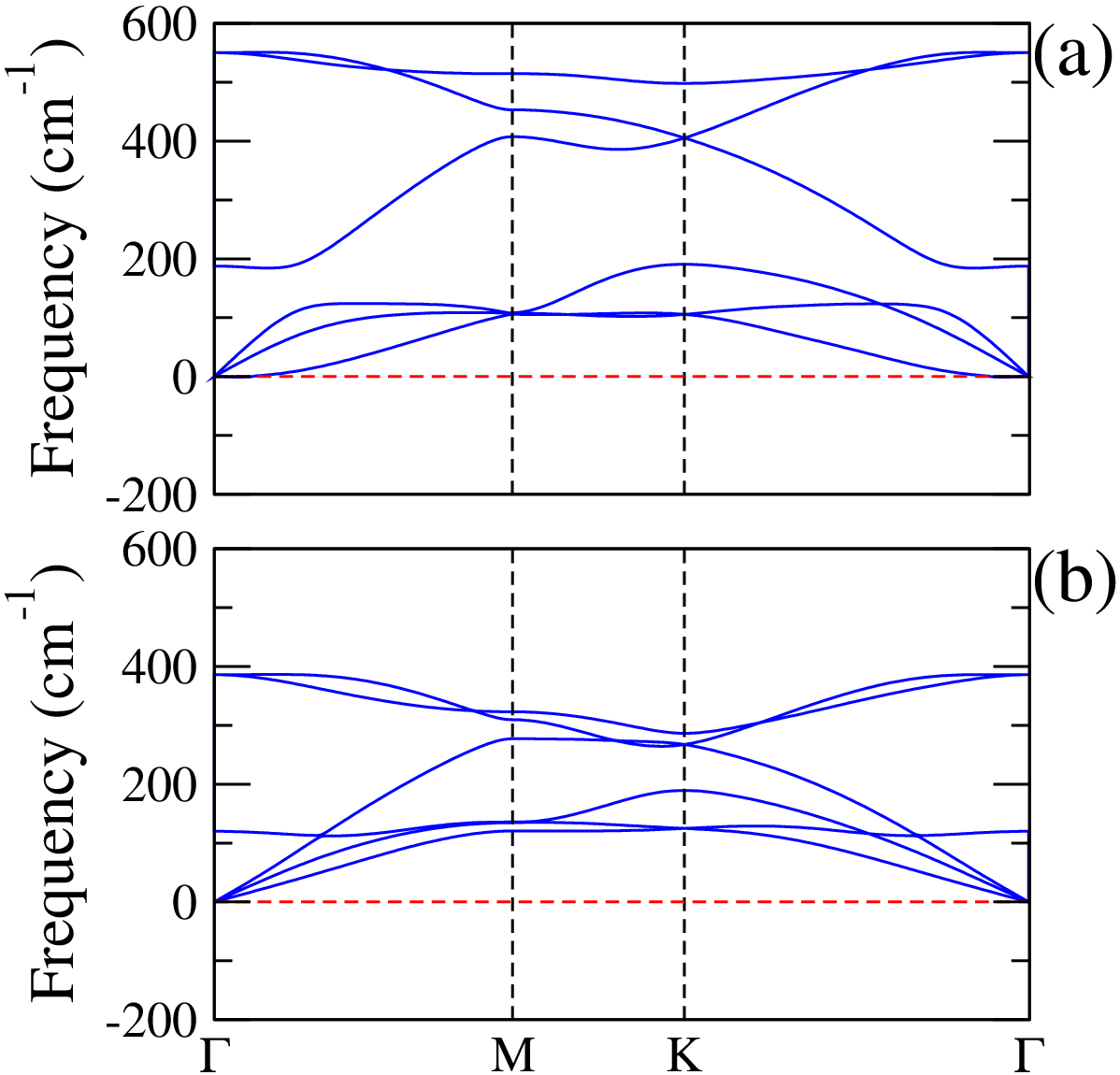}
\caption{Phonon spectrum of silicene (a) without strain and (b) under 10\%
biaxial tensile strain (courtesy of Ref.\ \cite{udo1}).}
\end{figure}

A quantum point contact of silicene has been put forward as high efficiency
spin filter with quantized conductance \cite{bansil}. 
In Fig.\ 3 the spin polarization is shown as a function of the
effective chemical potential $\mu_0$. For positive $\mu_0<0.1$ eV
the current flows within the conduction band of the left valley with 98\% spin
down polarization, while for negative $\mu_0>-0.1$ eV the current 
flows within the valence band of the right valley with 98\% spin up polarization.
By changing the effective chemical potential $\mu_0$ by means of a gate, the spin
polarization can be reversed. The authors have also demonstrated that a transition from
an indirect gap semiconductor to a zero gap state can be achieved by a Zeeman field.

\begin{figure}[t]
\includegraphics[width=0.4\textwidth,clip]{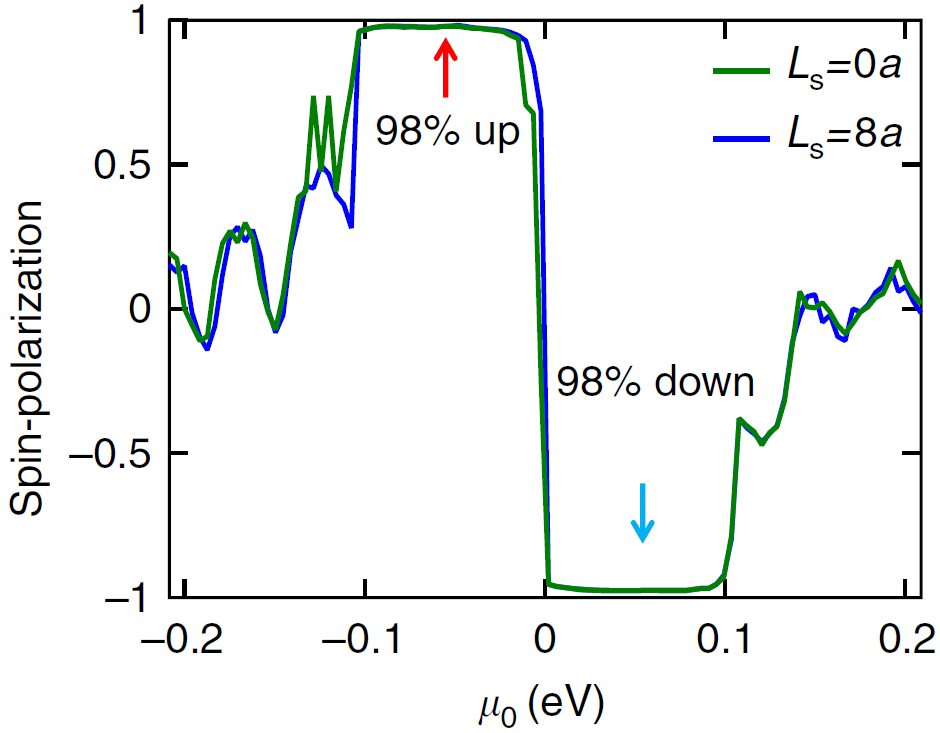}
\caption{Spin polarization versus effective chemical potential (courtesy of Ref.\ \cite{bansil}).}
\end{figure}

An external perpendicular electric field $E_z$ applied to silicene breaks the
sublattice symmetry due to different onsite energies, thus opening a band gap $\Delta$
at the K/K$'$-points. This is not possible in graphene, because the two sublattices
lie in one plane. The variation of the band gap as a function of $E_z$ has been
calculated in Ref.\ \cite{Drummond} using the local density approximation and
generalized gradient approximation. However, the spin orbit coupling has not been
included in these calculations though it plays a key role in the vicinity of the
K/K$'$-points. In particular, it leads to the creation of massive Dirac particles
and significantly modifies the Fermi velocity \cite{n11}.
Quantum transport calculations for a dual-gated silicene field
effect transistor, see Fig.\ 4(a) for a schematic view, accordingly find under $E_z$ a
transport gap \cite{Ni}. As compared to a single-gated field effect transistor, here
both the doping level and $E_z$ can be controlled. Since a $h$-BN trilayer remains
semiconducting upto $E_z=1.0$ V/\AA, it can be utilized as buffer to
prevent tunneling between the silicene sheet and gate. The effect of $E_z$ on the
transmission coefficient and projected density of states is addressed in Fig.\ 4(b,c).
A transport gap of 0.14 eV for $E_z=1.0$ eV indicates a strong dependence on the
electric field. The transmission eigenstates are shown in Fig.\ 4(d). Eigenvalues
of 0.09 for the OFF state and 0.87 for the ON state are obtained, reflecting a
remarkable OFF/ON ratio.

\begin{figure}[t]
\includegraphics[width=0.48\textwidth,clip]{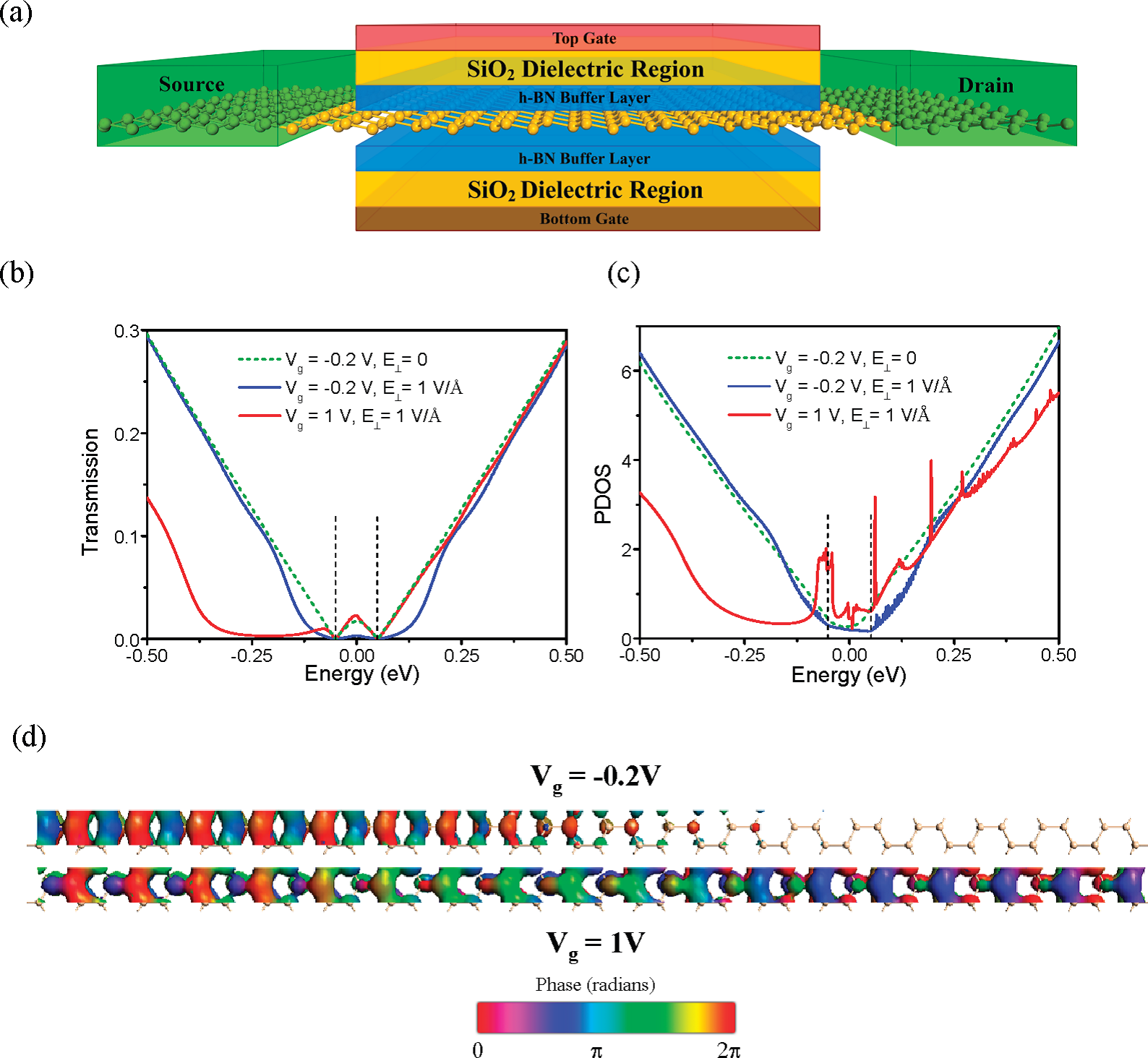}
\caption{Dual-gated silicene field effect transistor: (a) schematic view,
(b) transmission spectrum, (c) projected density of states and (d) transmission
eigenstates for the OFF and ON states (courtesy of Ref.\ \cite{Ni}).}
\end{figure} 

The strong spin orbit coupling in silicene, as compared to graphene, combined with an
external electric field can give rise to special phase transitions \cite{kaloni-sci}.
While we have a topologically nontrivial insulator for $E_z=0$ meV, spin splitting
evolves and the band gap thus starts to shrink when $E_z$ is switched on. The
band gap closes for $E_z=3.6$ meV and grows again in an almost linear manner when
$E_z$ is further enhanced, giving rise to a band insulator. These effects could not
be observed in Refs.\ \cite{Drummond,Ni}, because the spin orbit coupling was not
taken into account.

\subsection{Interaction with substrates} 
Silicene can be grown on Ag(111) \cite{Quaresima,Chen,n1}, but theory
predicts that the Dirac cone will disappear \cite{prl13}. The band structure becomes
complicated with the states at the Fermi energy clearly not being due to the Si $3p_z$
orbitals. Instead, there is strong hybridization between the Si and Ag orbitals, which
explains why the electronic behaviour changes fundamentally. This picture has been
confirmed experimentally in Ref.\ \cite{afm14}. For overcoming the hybridization issue
and keeping the Dirac cone intact, the interaction with the substrate has to be reduced.
This can be achieved by semiconducting substrates, which are also promising for
device applications similar to those already demonstrated for graphene on $h$-BN and
SiC(0001) \cite{Young1,Dimitrakopoulo,Seyller,Gannett,Santoso,xue,Watanabe2,Zhao3}.
 
The structural and electronic properties of superlattices of silicene and $h$-BN
have been addressed in Ref.\ \cite{kaloni-sci}. By the large band gap of $h$-BN,
it is not expected that B or N states appear in the vicinity of the Fermi energy.
Indeed, a cone with a 1.6 meV band gap (as a consequence of the spin orbit coupling) is
observed, which traces back to the Si $3p_z$ orbitals and is slightly hole doped due
to a small charge transfer towards the substrate. Some results for
SiC(0001) and $h$-BN substrates are summarized in Table II. It is found that
silicene behaves as a free-standing sheet with a distance of 3.0 \AA\ to the substrate.
The cohesive energy is less than 90 meV per Si atom. Moreover, the Dirac point remains
at the Fermi energy on H-passivated Si-SiC(0001), see Fig.\ 5, whereas on
H-passivated C-SiC(0001) metallicity is induced by charge transfer towards the silicene
sheet. This effect is understood in terms of the work function, since the valence band
maximum of H-passivated C-SiC(0001) lies above that of silicene. Stability of silicene
also has been predicted on Cl-passivated Si(111) as well as on CaF$_2$(111) \cite{n4}.
Interestingly, Ca-intercalation of multilayer silicene results in a Dirac cone
(shifted substantially in energy due to charge transfer from the Ca atoms), which
gives rise to a promising alternative from the preparation point of view \cite{n5}.

\begin{table*}[t]
\begin{tabular}{c||c|c|c|c|c}
                & Silicene cell &Substrate Cell& Lattice mismatch (\%) & Si-Si bond length (\AA) & Binding energy (meV/atom)\\
\hline\hline
silicene/$h$-BN &$4\times4\times1$ &$6\times6\times1$ &2.0 &2.27-2.28 & 89\\
\hline
silicene/Si-SiC(0001) &$4\times4\times1$& $5\times5\times1$ &0.1 &2.27-2.28 & 67\\
\hline
silicene/C-SiC(0001)  &$4\times4\times1$& $5\times5\times1$ &0.1 &2.27-2.28 & 84\\
\end{tabular}
\caption{Silicene on $h$-BN, Si-SiC(0001), and C-SiC(0001) substrates (courtesy of Ref.\ \cite{Liu1}).}
\end{table*}

\begin{figure}[t]
\includegraphics[width=0.48\textwidth,clip]{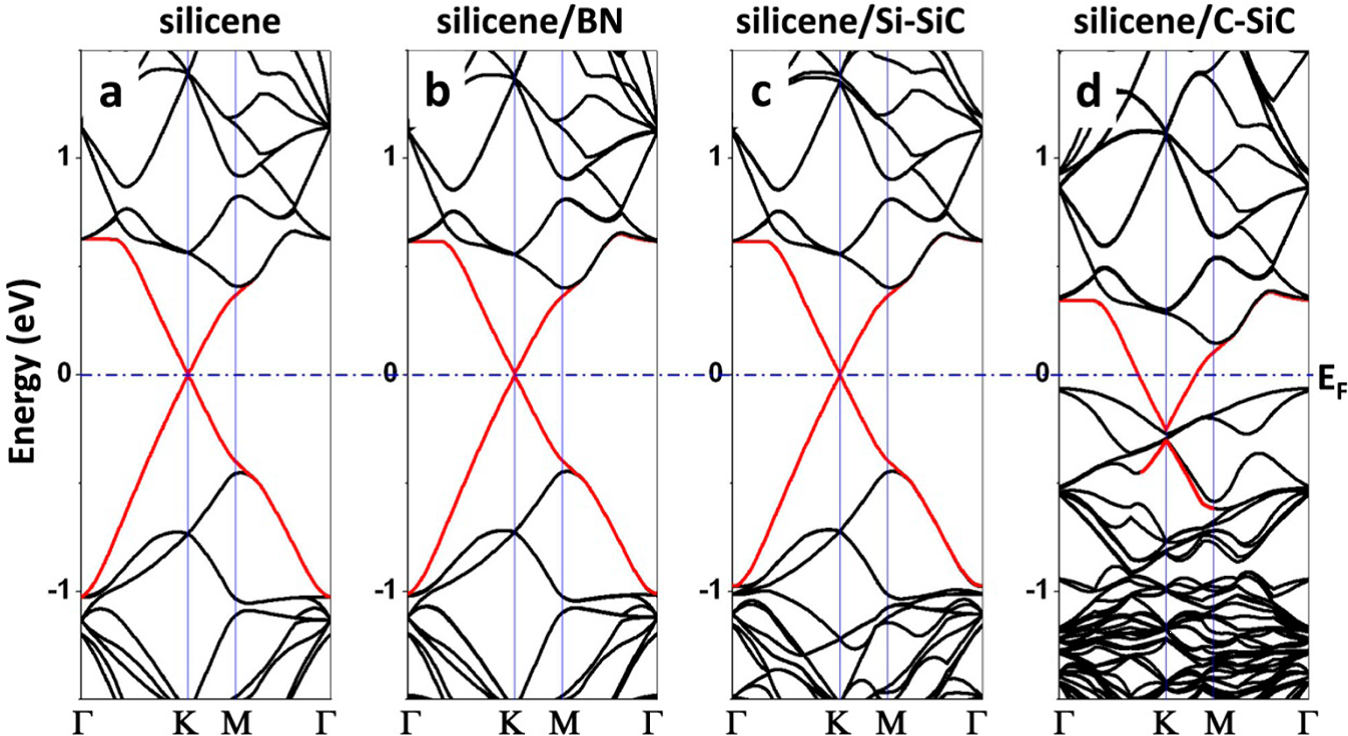}
\caption{Band structures of (a) pristine silicene, (b) silicene on $h$-BN, (c)
silicene on Si-SiC(0001), and (d) silicene on C-SiC(0001). The red lines highlight
the $\pi$ and $\pi^*$ bands (courtesy of Ref.\ \cite{Liu1}).} 
\end{figure}

Silicene on MoS$_2$ substrate shows a Dirac cone with a 70 meV band gap (much larger
than graphene on MoS$_2$) and therefore is interesting for electronic 
devices \cite{pccp14}. In the energetically favorable structure one Si
sublattice is located above S atoms and the other one above hollow sites of MoS$_2$. 
The binding energy of 120 meV per Si atom is close to that of silicene on GaS
(126 meV) \cite{apl13} but significantly larger than reported for $h$-BN
and SiC(0001), see Table II. The carrier density, effective mass, and mobility are
hardly influenced by the presence of MoS$_2$ \cite{Lee,Li12,Hiram,sant1,sant2}.
Accordingly, the bands at the Fermi energy have $\pi$ and $\pi^*$ character.
Si has been deposited on MoS$_2$ by molecular beam epitaxial growth and the local
structural and electronic properties have been investigated by scanning tunneling
microscopy and spectroscopy \cite{Chiappe}. The experiments show 2D domains of
Si with hexagonal atomic arrangement but unexpected electronic character, which can
be interpreted as a new allotropic form of Si.
We note that the interaction of silicene with semiconducting substrates also can be
quite substantial, e.g., metallic or magnetic states can be achieved \cite{DAS}. 
For AlAs(111), AlP(111), and GaAs(111) the surface magnetism of the substrate is
suppressed when silicene is attached, while for GaP(111) and ZnSe(111) it is enhanced.

A silicene field effect transistor with room temperature mobility of about
100 cm$^2$V$^{-1}$s$^{-1}$ has been demonstrated in Ref.\ \cite{Li-Tao}. The
fabrication process resulting in a silicene channel with Ag electrodes is shown in
Fig.\ 6. After epitaxial growth of silicene on Ag(111), an Al$_2$O$_3$ capping layer
is applied in situ, the structure is detached from the mica substrate, turned over,
and transferred onto a SiO$_2$ substrate. Then an etch-back approach is used to
define the source and drain contacts in the Ag film. In the region where the Ag has
been removed silicene is only in contact with semiconducting Al$_2$O$_3$ (weak
interaction). Electrical measurements at
room temperature demonstrate a transport behaviour similar to that of graphene.
A linear dependence of the drain current on the drain voltage confirms an Ohmic
contact. The fact that the silicene channel loses its Raman and electrical signatures
while exposed to air is likely to reflect degradation to an amorphous insulator and
indicates that the observed characteristics indeed are due to the silicene channel.

\begin{figure}[t]
\includegraphics[width=0.48\textwidth,clip]{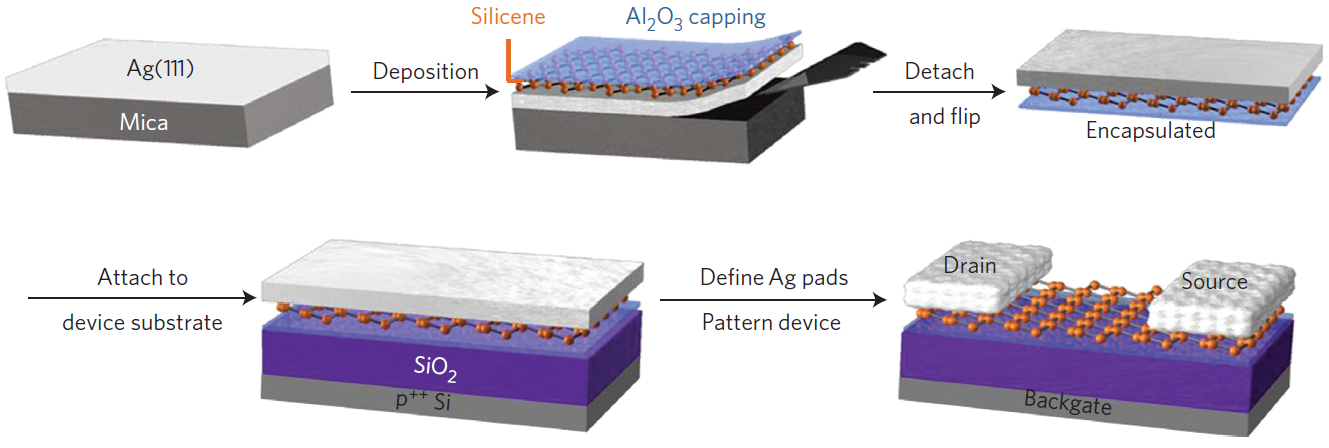}
\caption{Schematic fabrication process of a silicene field effect transistor
(courtesy of Ref.\ \cite{Li-Tao}).}
\end{figure} 

Epitaxial growth of silicene on ZrB$_2$(0001) has been reported in Ref.\ \cite{Fleurence}.
Fig.\ 7(a) shows a large-scale image from scanning tunneling microscopy with domain
boundaries running along the $\langle11\bar20\rangle$ direction. The
zoom in Fig.\ 7(b) highlights the honeycomb geometry with a lattice
constant of 3.65 \AA. The buckling is enhanced and loses its regularity due to
strong interaction with the substrate, resulting in a large band
gap of 250 meV, while the Si-Si bond length is maintained. Silicene on Ir(111)
after annealing at 670 K forms a $\sqrt7\times\sqrt7$ superstructure with respect
to the substrate, as determined by low energy electron diffraction \cite{Meng}.

\begin{figure}[t]
\includegraphics[width=0.35\textwidth,clip]{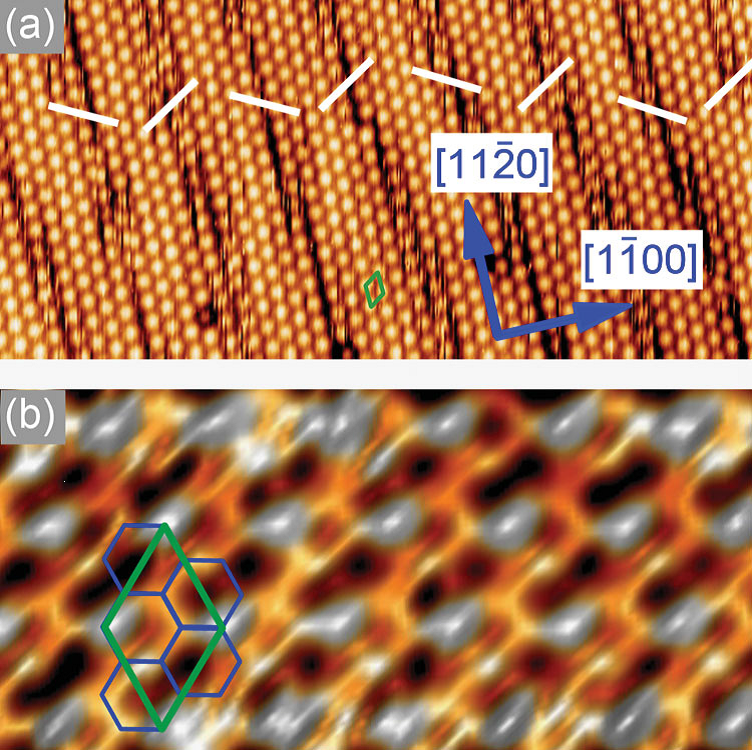}
\caption{Scanning tunneling microscopy images of silicene on ZrB$_2$(0001): (a) Large-scale
and (b) zoomed view (courtesy of Ref.\ \cite{Fleurence}).}
\end{figure}

\subsection{Functionalization}
Tuning of the band gap can be achieved by functionalization
\cite{Houssa-apl,Biberiana,Osborn1,Zheng2}, similar to graphene
\cite{Ramanathan1,Xian,Robinson,Shenoy,Balakrishnan,Wagner}. Note that in this respect
theoretical results obtained for free-standing silicene will also apply in the
presence of a weakly interacting substrate. A sizable band gap
without disturbing the electronic characteristics of silicene is particularly
fruitful for field effect transistors \cite{Li-Tao}. The authors of Ref.\ \cite{Fei}
have proposed that this goal is achieved by decoration with Li, Na, K, Rb,
and Cs atoms (growing charge transfer along the series). The binding energy is found
to decrease for increasing coverage due to dopant-dopant interaction.

\begin{figure}[b]
\includegraphics[width=0.48\textwidth,clip]{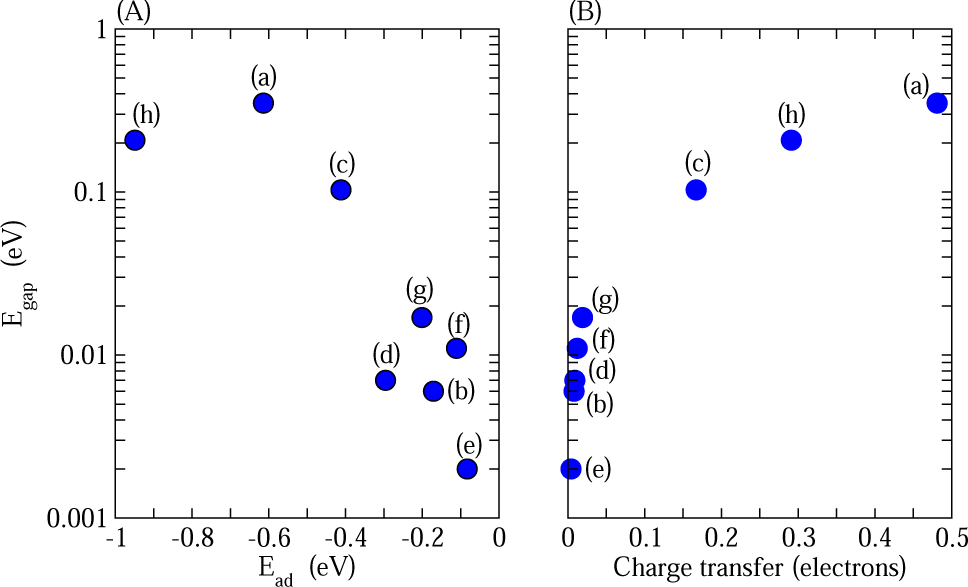}
\caption{Band gap as a function of the (left) binding energy and (right) charge 
transfer for (a) acetone, (b) acetonitrile, (c) ammonia, (d) benzene, (e) methane,
(f) methanol, (g) ethanol, and (h) toluene adsorption on silicene (courtesy
of Ref.\ \cite{freund}).}
\end{figure}

Adsorption of small organic molecules is also a very efficient way
to modify the band gap \cite{freund}. The molecules induce a perpendicular electric
field, which breaks the sublattice symmetry and thus opens a band gap
\cite{Ramasubramaniam}. The adsorption energy varies from 110 meV for
acetone to 905 meV for toluene, see Fig.\ 8(left). These values are higher than found
for the adsorption of similar molecules on graphene \cite{JACS}, which indicates
that silicene is a good host. Band gaps between 6 meV (acetonitrile) and 350 meV
(acetone) have been reported as well as mobilities even higher than in graphene, which is
important for electronic devices. The charge transfer from/to the adsorbed molecule,
see Fig.\ 8(right), is essentially proportional to the size of the band gap.
The effect of H, CH$_3$, OH, and F decoration on the electronic properties of monolayer
and bilayer silicene has been studied in Ref.\ \cite{Denis}. As compared to graphene,
the bonds with Si atoms are substantially weaker, the strongest being the Si-F bond
with a binding energy of 4.99 eV per atom. H and CH$_3$ attached to monolayer silicene
provide optical band gaps of 3.2 eV, while OH and F result in metallicity. F also
strongly modifies the electronic structure of bilayer silicene and yields a large
band gap of 3.0 eV, whereas for the other adsorbants band gaps between 0.1 eV and
0.6 eV are found. 

A rich phase diagram of silicene (and analogously of germanene) in the presence of
exchange and perpendicular electric fields results already from a simplified model
Hamiltonian, comprising quantum spin Hall, quantum anomalous Hall, band insulating,
and valley-polarized metallic phases \cite{ezawa}. The quantum spin Hall state has
been confirmed by calculating the $Z_2$ topological invariant from the first principles
electronic band structure of silicene \cite{n6} and by a tight-binding approach for
germanene nanoribbons \cite{n7}. 
A quantum anomalous Hall state has been described theoretically for
Co-decorated silicene \cite{kaloni-prb}, while for other transition metals (Ti, V, Cr,
Mn, and Fe) the state is prohibited by pertubations of the Dirac cone. In addition, it
has been demonstrated that only for a sufficiently
low Co coverage the quantum anomalous Hall state is realized. For larger Co coverage
the Co-Co interaction modifies the shape of the electronic bands strongly in the
vicinity of the Fermi energy. In contrast to Ref.\ \cite{kaloni-prb}, topologically
non-trivial states have been predicted for V-decorated silicene in Ref.\
\cite{Zhang-sci}, but only if an onsite Coulomb interaction is taken into account.
The quantum anomalous Hall state is prohibited in the case of Ti, Cr, and Mn decoration
by hybridization between the Si $3p_z$ and transition metal $3d$ states.

\section{Germanene}

\begin{figure}[b]
\includegraphics[width=0.48\textwidth,clip]{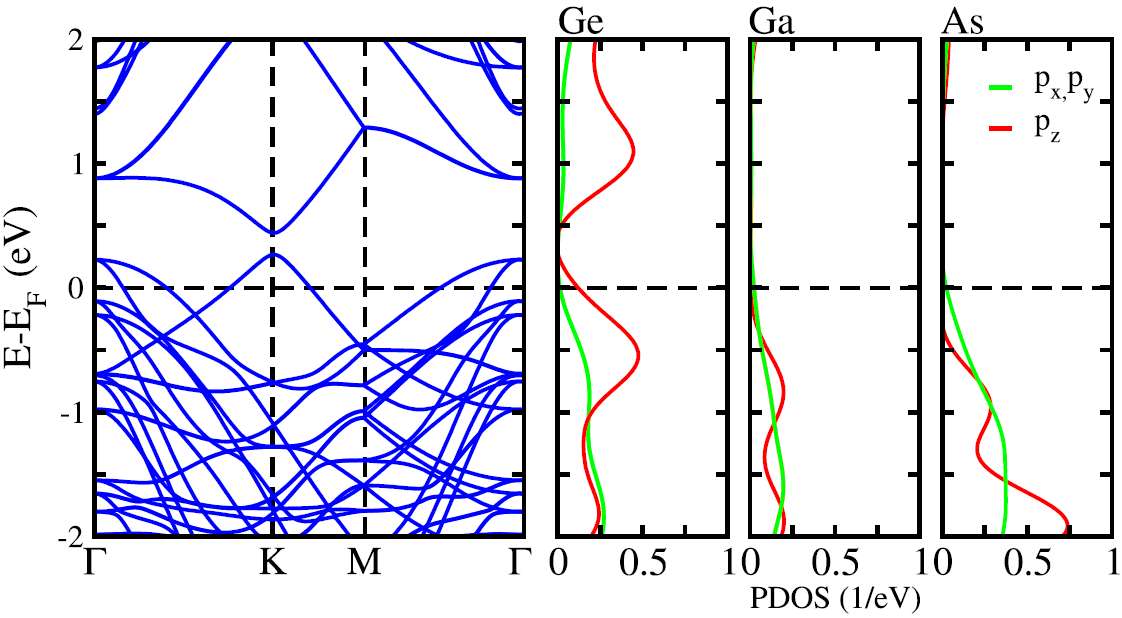}
\caption{Band structure of germanene on GaAs(0001) with partial densities
of states (courtesy of Ref.\ \cite{kaloni-jap}).}
\end{figure}

Semiconducting GaAs(0001) interacts strongly with germanene, forming bonds with a
binding energy of 568 meV per Ge atom, while on H-passivated GaAs(0001)
the interaction is reduced to 69 meV per Ge atom \cite{kaloni-jap}. 
This value comes close to the experimental binding energy of graphene sheets 
in graphite \cite{Ulbricht}. According to Fig.\ 9, the electronic bands are forming
hole pockets at the $\Gamma$-point (due to Ga and As states), while a split
Dirac cone (energy gap of 175 meV, as a consequence of the broken sublattice symmetry)
appears above the Fermi energy. Application of an appropriate voltage can bring it back
to the Fermi energy to induce a semiconducting state.

It has been shown that germanene can be grown on Au(111) with a binding energy of 110 meV
per Ge atom \cite{NJP}. Since this value is lower than the binding energy in bulk Ge,
formation of Ge clusters is to be expected. Moreover, ab-initio simulations point to a
honeycomb lattice with weak corrugations. Single phase growth of germanene has been
reported on Al(111) \cite{nano-2015}. The energetics and electronic structure
suggest that graphene also can be used as substrate \cite{PRB-germanene}. The
interaction is weak and a band gap of 57 meV is realized, which is more than twice
that of the pristine material (24 meV). Germanene on ZnSe(0001) shows an indirect band
gap of 400 meV, whereas a charge transfer analysis points to only weak interaction
\cite{ass}. An electric field of $E_z=0.6$ V/\AA\ results in transition to a direct
band gap.

Concerning metallic substrates, germanene has been claimed to be achieved on Pt(111)
by deposition of Ge atoms with subsequent annealing \cite{Li,Bampoulis}. Low energy
electron diffraction and scanning tunneling microscopy measurements have been
interpreted in terms of commensurability with a $\sqrt19\times\sqrt19$
pattern. These findings have been disputed by $\check{\rm S}$vec and co-workers, who
rule out the formation of silicene or germanene on transition metal surfaces
\cite{Hapala}. Based on various experimental and theoretical techniques, their
comprehensive study instead points to the formation of an ordered 2D surface alloy.
On the other hand, deposition of Pt on Ge(110)
followed by annealing at 1100 K results in three-dimensional Pt-Ge nanocrystals
\cite{Bampoulis}. Interestingly, the outermost layer of these crystals possesses
a honeycomb structure, consisting of two hexagonal sublattices displaced by about
0.2 \AA\ with respect to each other. The nearest neighbor distance of 2.5 \AA\
agrees well with the Ge-Ge bond length in pristine germanene.

Alkali metal atoms are found to be strongly bound on germanene and capable of
engineering a band gap of 20 meV to 310 meV \cite{physicae}. The observed large
carrier mobility has been put forward for high speed field effect transistors. The
structural and electronic properties induced by alkali, alkaline-earth, group-III, and
transition metal atoms adsorbed on germanene have been studied in Ref.\ \cite{pccp-14}.
Alkali metals (Li, Na, and K) become ionized and lead to metallicity, while
alkali-earth metals (Be, Mg, and Ca) result in a mixed ionic/covalent bonding
and a semiconducting state. For group-III atoms (Al, Ga, and In) the interaction is
enhanced (mixed ionic/covalent bonding) and transition metals (Ti, V, Cr, Fe, Co, and
Ni) finally result in large distortions of the germanene sheet, where metallic,
half-metallic, and semiconducting states can be realized. The effect of
Mn decoration has been investigated in Ref.\ \cite{kaloni-jpcc} as a function of
the coverage.

\vfill

\section{Summary}
Substantial progress has been achieved towards applications of silicene and germanene.
Silicene nowadays can be grown on different Ag surfaces, Ir(111), MoS$_2$, and ZrB$_2$(0001).
Energetical stability also has been predicted on $h$-BN, H-passivated
C-SiC(0001) and Si-SiC(0001), Cl-passivated Si(111), CaF$_2$, AlAs(111), AlP(111),
GaAs(111), GaP(111), ZnSe(111), and ZnS(111). Particularly the semiconducting
substrates are of technological interest, because the interaction with silicene is
reduced and the Dirac states consequently are maintained, even though a sizable band
gap may develop. In the cases of $h$-BN and H-passivated Si-SiC(0001) the effects of
the interaction are close to negligible so that
quasi-free-standing silicene may be realized. Band gap
control is achieved by external electric fields and decoration with alkali,
alkali-earth, and transition metal atoms. According to theoretical predictions, the
decoration approach furthermore provides access to a multitude of exotic phases.
Germanene has been grown on Al(111) and Au(111), and the possibility of
growth on GaAs(0001), graphene, and ZnSe(0001) has been put forward. Again
the band gap is susceptible to engineering approaches based on atomic decoration.

\begin{acknowledgments}
Research reported in this publication was supported by the King Abdullah University of
Science and Technology (KAUST). G.S.\ acknowledges funding
from the Natural Sciences and Engineering Council of Canada (NSERC, Discovery Grant).
M.S.F.\ acknowledges support by the Natural Sciences and Engineering Research Council
(NSERC) of Canada, the Canada Research Chair program, Canada Foundation for Innovation
(CFI), the Manitoba Research and Innovation Fund, and the University of Manitoba.
\end{acknowledgments}


\begin{thebibliography}{520}
\bibitem{Novoselov}K. S. Novoselov, A. K. Geim, S. Morozov, D. Jiang, Y. Zhang, S. V. Dubonos, I. V. Grigorieva, and A. A. Firsov, Science {\bf 306}, 666 (2004).

\bibitem{Xu}M. Xu, T. Liang, M. Shi, and H. Chen, Chem. Rev. {\bf 113}, 3766 (2013).

\bibitem{Watanabe}K. Watanabe, T. Taniguchi, and H. Kanda, Nat. Mater. {\bf 3}, 404 (2004).

\bibitem{Kubota}Y. Kubota, K. Watanabe, O. Tsuda, and T. Taniguchi, Science {\bf 317}, 932 (2007).

\bibitem{Meyer}D. Pacil\'e, J. C. Meyer, \c{C}. \"O. Girit, and A. Zettl, Appl. Phys. Lett. {\bf 92}, 133107 (2008).

\bibitem{Novoselov3}K. S. Novoselov, V. I. Fal'ko, L. Colombo, P. R. Gellert, M. G. Schwab, and K. Kim, Nature {\bf 490}, 192 (2012).

\bibitem{Geim2} A. K. Geim, Nature {\bf 499}, 419 (2013).

\bibitem{sant1}K. C. Santosh, R. L. Pazos, R. Addou, R. Wallace, and K. Cho, Nanotechnology \textbf{25}, 375703 (2014).

\bibitem{sant2}A. Azcatl, S. McDonnell, K. C. Santosh, X. Peng, H. Dong, X. Qin, R. Addou, G. Mordi, N. Lu, J. Kim, M. J. Kim, K. Cho, and R. M. Wallace, Appl. Phys. Lett. \textbf{104}, 111601 (2014).

\bibitem{sahin1}S. Cahangirov, M. Topsakal, E. Akt\"urk, H. \c{S}ahin, and S. Ciraci, Phys. Rev. Lett. {\bf 102}, 236804 (2012).

\bibitem{Fleurence}A. Fleurence, R. Friedlein, T. Ozaki, H. Kawai, Y. Wang, and Y. Yamada-Takamura, Phys. Rev. Lett. {\bf 108}, 245501 (2012).

\bibitem{Quaresima} P. Vogt, P. D. Padova, C. Quaresima, J. Avila, E. Frantzeskakis, M. C. Asensio, A. Resta, B. Ealet, and G. L. Lay, Phys. Rev. Lett. {\bf 108}, 155501 (2012).

\bibitem{Chen}L. Chen, C.-C. Liu, B. Feng, X. He, P. Cheng, Z. Ding, S. Meng, Y. Yao, and K. Wu, Phys. Rev. Lett. {\bf 109}, 056804 (2012). 

\bibitem{Meng}L. Meng, Y. Wang, L. Zhang, S. Du, R. Wu, L. Li, Y. Zhang, G. Li, H. Zhou, W. A. Hofer, and H.-J. Gao, Nano Lett.  {\bf 13}, 685 (2013).

\bibitem{datta}D. Jose and A. Datta, Acc. Chem. Res. \textbf{47}, 593 (2014).

\bibitem{kaloni-jap}T. P. Kaloni and U. Schwingenschl\"ogl, J. Appl. Phys. {\bf 114}, 184307 (2013).

\bibitem{Lia}L. Lia and M. Zhao, Phys. Chem. Chem. Phys. {\bf 15}, 16853 (2013).

\bibitem{NJP}M. E. D\'avila, L. Xian, S. Cahangirov, A. Rubio, and G. Le Lay, New J. Phys. \textbf{16}, 095002 (2014).

\bibitem{nano-2015}M. Derivaz, D. Dentel, R. Stephan, M.-C. Hanf, A. Mehdaoui, P. Sonnet, and C. Pirri, Nano Lett. \textbf{15}, 2510 (2015).

\bibitem{Ongun}V. O. \"Oz\c{c}elik, S. Cahangirov, and S. Ciraci, Phys. Rev. Lett. {\bf 112}, 246803 (2014).

\bibitem{Takeda}K. Takeda and K. Shiraishi, Phys. Rev. B {\bf 50}, 14916 (1994).

\bibitem{Voon}G. G. Guzm\'an-Verri and L. C. L. Y. Voon, Phys. Rev. B {\bf 76}, 075131 (2007).

\bibitem{ciraci3}S. S. Cahangirov, M. Topsakal, E. Akt\"urk, H. \c{S}ahin, and S. Ciraci, Phys. Rev. Lett. {\bf 102}, 236804 (2009).

\bibitem{n1}W. Ke-Hui, Chin. Phys. B \textbf{24}, 086802 (2015).

\bibitem{Lin2}C.-L. Lin, R. Arafune, K. Kawahara, M. Kanno, N. Tsukahara, E. Minamitani, Y. Kim, M. Kawai, and N. Takagi, Phys. Rev. Lett. {\bf 110}, 076801 (2013).

\bibitem{Rubio}S. Cahangirov, M. Audiffred, P. Tang, A. Iacomino, W. Duan, G. Merino, and A. Rubio, Phys. Rev. B {\bf 88}, 035432 (2013).

\bibitem{Wang}Y.-P. Wang, and H.-P. Cheng, Phys. Rev. B {\bf 87}, 245430 (2013).

\bibitem{kaloni-sci} T. P. Kaloni, M. Tahir, and U. Schwingenschl\"ogl, Sci. Rep. {\bf 3}, 3192 (2013).

\bibitem{Liu1}H. Liu, J. Gao, and J. Zhao, J. Phys. Chem. C {\bf 117}, 10353 (2013).

\bibitem{Quhe} R. Quhe, Y. Yuan, J. Zheng, Y. Wang, Z. Ni, J. Shi, D. Yu, J. Yang, and J. Lu, Sci. Rep. {\bf 4}, 5476 (2014).

\bibitem{Lebegue}S. Lebegue and O. Eriksson, Phys. Rev. B {\bf 79}, 115409 (2009).

\bibitem{liu-epl}G. Liu, M. S. Wu, C. Y. Ouyang, and B. Xu, EPL {\bf 99}, 17010 (2012).

\bibitem{n11}L. Matthes, O. Pulci, and F. Bechstedt, J. Phys.: Condens. Matter \textbf{25}, 395305 (2013).

\bibitem{udo1}T. P. Kaloni, Y. C. Cheng, and U. Schwingenschl\"ogl, J. Appl. Phys. {\bf 113}, 104305 (2013).

\bibitem{Dresselhaus1} B. Wang, J. Wu, X. Gu, H. Yin, Y. Wei, R. Yang, and M. Dresselhaus, Appl. Phys. Lett. {\bf 104}, 081902 (2014).

\bibitem{Drummond}N. D. Drummond, V. Z\'olyomi, and V. I. Fal'ko, Phys. Rev. B {\bf 85}, 075423 (2012).

\bibitem{Ni} Z. Ni, Q. Liu, K. Tang, J. Zheng, J. Zhou, R. Qin, Z. Gao, D. Yu, and J. Lu, Nano Lett. {\bf 12}, 113 (2012).

\bibitem{prb-ezawa}M. Ezawa, Phys. Rev. B \textbf{86}, 161407 (2012).

\bibitem{iop}H. H. G\"urel, V. O. \"Ozcelik, and S. Ciraci, J. Phys.: Condens. Matter \textbf{25}, 305007 (2013).

\bibitem{pccp} Y.-C. Zhao and J. Ni, Phys. Chem. Chem. Phys. \textbf{16}, 15477 (2014).

\bibitem{Houssa-apl} M. Houssa, E. Scalise, K. Sankaran, G. Pourtois, V. V. Afanasev, and A. Stesmans, Appl. Phys. Lett. {\bf 98}, 223107 (2011).

\bibitem{jacob-prb} W. Wei and T. Jacob, Phys. Rev. B {\bf 88}, 045203 (2013).
 
\bibitem{Hussain-pccp}T. Hussain, T. Kaewmaraya, S. Chakraborty, and R. Ahuja, Phys. Chem. Chem. Phys. {\bf 21}, 18900 (2013).

\bibitem{Ouyang-epl}G. Liu, X. L. Lei, M. S. Wu, B. Xu, and C. Y. Ouyang, EPL {\bf 106}, 47001 (2014).

\bibitem{freund}T. P. Kaloni, G. Schreckenbach, and M. S. Freund, J. Phys. Chem. C \textbf{118}, 23361 (2014).

\bibitem{Fei}R. Quhe, R. Fei, Q. Liu, J. Zheng, H. Li, C. Xu, Z. Ni, Y. Wang, D. Yu, Z. Gao, and J. Lu, Sci. Rep. {\bf 2}, 853 (2012).

\bibitem{kaloni-rrl}T. P. Kaloni and U. Schwingenschl\"ogl, Phys. Status Solidi RRL \textbf{8}, 685 (2014).

\bibitem{Lin-prb} X. Lin and J. Ni, Phys. Rev. B {\bf 86}, 075440 (2012).

\bibitem{Zhao-prb}J. Zhang, B. Zhao, and Z. Yang, Phys. Rev. B {\bf 88}, 165422 (2013).

\bibitem{Zhang-sci} X.-L. Zhang, L.-F. Liu, and W.-M. Liu, Sci. Rep. {\bf 3}, 2908 (2013).

\bibitem{kaloni-prb14}T. P. Kaloni, N. Singh, and U. Schwingenschl\"ogl, Phys. Rev. B {\bf 89}, 035409 (2014).

\bibitem{n8}L. Matthes, P. Gori, O. Pulci, and F. Bechstedt, Phys. Rev. B \textbf{87}, 035438 (2013).

\bibitem{n9}W. Wei, Y. Dai, B. Huang, and T. Jacob, Phys. Chem. Chem. Phys. \textbf{15}, 8789 (2013).

\bibitem{n10}O. Pulci, P. Gori, M. Marsili, V. Garbuio, R. Del Sole, and F. Bechstedt, EPL \textbf{98}, 37004 (2012).

\bibitem{Cai-prb} Y. Cai, C.-P. Chuu, C. M. Wei, and M. Y. Chou, Phys. Rev. B {\bf 88}, 245408 (2013).

\bibitem{kaloni-cpl} T. P. Kaloni and U. Schwingenschl\"ogl, Chem. Phys. Lett. {\bf 583}, 137 (2013).

\bibitem{Stein-prb}J.-A. Yan, R. Stein, D. M. Schaefer, X.-Q. Wang, and M. Y. Chou, Phys. Rev. B \textbf{88}, 121403 (2013) 

\bibitem{Roome}N. J. Roome and J. D. Carey, ACS Appl. Mater. Interfaces {\bf 6}, 7743 (2014).

\bibitem{ma1}Y. Ma , Y. Dai, C. Niu, and B. Huang, J. Mater. Chem. {\bf 22}, 12587 (2012).

\bibitem{Kusmartsev}A. O'Hare, F. V. Kusmartsev, and K. I. Kugel, Nano Lett. {\bf 12}, 1045 (2012).

\bibitem{Chen-jpc}Y. Li and Z. Chen, J. Phys. Chem. C {\bf 118}, 1148 (2014)

\bibitem{si1}C. Si, J. Liu, Y. Xu, J. Wu, B.-L. Gu, and W. Duan, Phys. Rev. B {\bf 89}, 115429 (2014).

\bibitem{kaloni-jpcc} T. P. Kaloni, J. Phys. Chem. C \textbf{118}, 25200 (2014).

\bibitem{Zhang-pccp} S.-S. Li, C.-W. Zhang, W.-X. Ji, F. Li, P.-J. Wang, S.-J. Hu, S.-S. Yan, and Y.-S. Liu, Phys. Chem. Chem. Phys. 10.1039/C4CP01211A.

\bibitem{Dai-jmc}Y. Ma, Y. Dai, C. Niua, and B. Huanga, J. Mater. Chem. {\bf 22}, 12587 (2012).
 
\bibitem{Seixas}L. Seixas, J. E. Padilha, and A. Fazzio, Phys. Rev. B {\bf 89}, 195403 (2014).

\bibitem{Huang-prb}S.-M. Huang, S.-T. Lee, and C.-Y. Mou, Phys. Rev. B {\bf 89}, 195444 (2014).

\bibitem{Tabert}C. J. Tabert and E. J. Nicol, Phys. Rev. B {\bf 87}, 235426 (2013).

\bibitem{Ezawa-prb} S. Rachel and M. Ezawa, Phys. Rev. B {\bf 89}, 195303 (2014).

\bibitem{Yang-prb}K. Yang, S. Cahangirov, A. Cantarero, A. Rubio, and R. D'Agosta, Phys. Rev. B {\bf 89}, 125403 (2014). 

\bibitem{Zhao-pccp}L. Li and M. Zhao, Phys. Chem. Chem. Phys. {\bf 15}, 16853 (2013).

\bibitem{Bianco}E. Bianco, S. Butler, S. Jiang, O. D. Restrepo, W. Windl, and J. E. Goldberger, ACS Nano {\bf 7}, 4414 (2013).

\bibitem{n2}S. Cahangirov, V. O. \"Ozcelik, L. Xian, J. Avila, S. Cho, M. C. Asensio, S. Ciraci, and A. Rubio, Phys. Rev. B \textbf{90}, 035448 (2014).

\bibitem{n3}F. Matusalem, M. Marques, L. K. Teles, and F. Bechstedt, Phys. Rev. B \textbf{92}, 045436 (2015). 

\bibitem{xiao}X.-L. Wang, S. X. Dou, and C. Zhang, NPG Asia Materials \textbf{2}, 31 (2010).

\bibitem{peeters}H. Sahin and F. M. Peeters, Phys. Rev. B {\bf 87}, 085423 (2013).

\bibitem{neto}A. H. Castro Neto, F. Guinea, N. M. R. Peres, K. S. Novoselov, and A. K. Geim, Rev. Mod. Phys. {\bf 81}, 109 (2009).

\bibitem{Wehling}T. O. Wehling,  E. \c{S}a\c{s}\i o\u{g}lu, C. Friedrich, A. I. Lichtenstein, M. I. Katsnelson, and S. Bl\"ugel, Phys. Rev. Lett. {\bf 106}, 236805 (2011).

\bibitem{Scalise}E. Scalise, M. Houssa, G. Pourtois, B. van den Broek, V. Afanasev, and A. Stesmans, Nano Res. {\bf 6}, 19 (2013).%

\bibitem{Koskinen}P. Koskinen, S. Malola, and H. H\"akkinen, Phys. Rev. Lett. {\bf 101}, 115502 (2008).

\bibitem{Kwon}H. Shin, S. Kang, J. Koo, H. Lee, J. Kim, and Y. Kwon, J. Chem. Phys. {\bf 140}, 114702 (2014).

\bibitem{Jiang}C.-C. Liu, H. Jiang, and Y. Yao, Phys. Rev. B {\bf 84}, 195430 (2011).

\bibitem{PRB2013}M. Hu, X. Zhang, and D. Poulikakos, Phys. Rev. B {\bf 87}, 195417 (2013).

\bibitem{bansil}W.-F. Tsai, C.-Y. Huang, T.-R. Chang, H. Lin, H.-T. Jeng, and A. Bansil, Nat. Commun. {\bf 4}, 1500 (2013).

\bibitem{prl13} C.-L. Lin, R. Arafune, K. Kawahara, M. Kanno, N. Tsukahara, E. Minamitani, Y. Kim, M. Kawai, and N. Takagi, Phys. Rev. Lett. {\bf 110}, 076801 (2013).

\bibitem{afm14}N. W. Johnson, P. Vogt, A. Resta  P. D. Padova, I. Perez, D. Muir, E. Z. Kurmaev, G. Le Lay, and A. Moewes, Adv. Funct. Mater. \textbf{24}, 5253 (2014).

\bibitem{Young1}C. R. Dean, A. F. Young, I. Meric, C. Lee, L. Wang, S. Sorgenfrei, K. Watanabe, T. Taniguchi, P. Kim, K. L. Shepard, and J. Hone, Nat. Nanotechnol. {\bf 5}, 722 (2010).

\bibitem{Dimitrakopoulo} Y.-M. Lin, C. Dimitrakopoulos, K. A. Jenkins, D. B. Farmer, H.-Y. Chiu, A. Grill, and P. Avouris, Science {\bf 327}, 662 (2010).

\bibitem{Seyller}P. N. First, W. A. de Heer, T. Seyller, C. Berger, J. A. Stroscio, and J.-S. Moon, MRS Bull. {\bf 35}, 296 (2010).

\bibitem{Gannett}W. Gannett, W. Regan, K. Watanabe, T. Taniguchi, M. F. Crommie, and A. Zettl, Appl. Phys. Lett. {\bf 98}, 242105 (2011).

\bibitem{Santoso}S. L. Wong, H. Huang, Y. Wang, L. Cao, D. Qi, I. Santoso, W. Chen, and A. T. S. Wee, ACS Nano {\bf 5}, 7662 (2011).

\bibitem{xue}J. Xue, J. Sanchez-Yamagishi, D. Bulmash, P. Jacquod, A. Deshpande, K. Watanabe, T. Taniguchi, P. Jarillo-Herrero, and B. J. LeRoy, Nat. Mater. {\bf 10}, 282 (2011).

\bibitem{Watanabe2}M. Yankowitz, J. Xue, D. Cormode, J. D. Sanchez-Yamagishi,  K. Watanabe, T. Taniguchi, P. Jarillo-Herrero, P. Jacquod, and B. J. LeRoy, Nat. Phys. {\bf 8}, 382 (2012).

\bibitem{Zhao3}H. Liu, J. Gao, and J. Zhao, J. Phys.: Conf. Ser. {\bf 491}, 012007 (2014).

\bibitem{n4}S. Kokott, P. Pflugradt, L. Matthes, and F. Bechstedt, J. Phys.: Condens. Matter \textbf{26}, 185002 (2014).

\bibitem{n5}E. Noguchi, K. Sugawara, R. Yaokawa, T. Hitosugi, H. Nakano, and T. Takahashi, Adv. Mater. {\bf 27}, 856 (2015).   

\bibitem{pccp14}N. Gao, J. C. Li, and Q. Jiang, Phys. Chem. Chem. Phys. {\bf 16}, 11673 (2014).

\bibitem{apl13}Y. Ding and Y. Wang, Appl. Phys. Lett. {\bf 103}, 043114 (2013).

\bibitem{Lee}K. F. Mak, C. Lee, J. Hone, J. Shan, and T. F. Heinz, Phys. Rev. Lett. {\bf 105}, 136805 (2010).

\bibitem{Li12}Y. Li and Z. Chen, J. Phys. Chem. Lett. {\bf 4}, 269 (2012).

\bibitem{Hiram}J. H. Conley, B. Wang, J. I. Ziegler, R. F. Haglund, S. T. Pantelides, and K. I. Bolotin, Nano Lett. {\bf 13}, 3626 (2013).

\bibitem{Chiappe}D. Chiappe, E. Scalise, E. Cinquanta, C. Grazianetti, B. van den Broek, M. Fanciulli, M. Houssa, and A. Molle, Adv. Mater. \textbf{26}, 2096 (2014).   

\bibitem{DAS}A. Bhattacharya, S. Bhattacharya, and G. P. Das, Appl. Phys. Lett. {\bf 103}, 123113 (2013).

\bibitem{Li-Tao}L. Tao, E. Cinquanta, D. Chiappe, C. Grazianetti, M. Fanciulli, M. Dubey, A. Molle, and D. Akinwande, Nat. Nanotechnol. \textbf{10}, 227 (2015).

\bibitem{Biberiana}G. Le Lay, B. Aufray, C. L\'eandri, H. Oughaddou, J.-P. Biberian, P. D. Padova, M.E. D\'avila, B. Ealet, and A. Kara, Appl. Surf. Sci. {\bf 256}, 524 (2009).

\bibitem{Osborn1}T. H. Osborn, A. A. Farajian, O. V. Pupysheva, R. S. Aga, and L. C. L. Y. Voon, Chem. Phys. Lett. {\bf 511}, 101 (2011).

\bibitem{Zheng2}N. Gao, W. T. Zheng, and Q. Jiang, Phys. Chem. Chem. Phys. {\bf 14}, 257 (2012)

\bibitem{Ramanathan1}T. Ramanathan, A. A. Abdala, S. Stankovich, D. A. Dikin, M. Herrera-Alonso, R. D. Piner, D. H. Adamson, H. C. Schniepp, X. Chen, R. S. Ruoff, S. T. Nguyen, I. A. Aksay, R. K. Prud'Homme, and L. C. Brinson, Nat. Nanotechnol. {\bf 3}, 327 (2008).

\bibitem{Xian}J.-A. Yan, L. Xian, and M. Y. Chou, Phys. Rev. Lett. {\bf 103}, 086802 (2009).

\bibitem{Robinson}J. T. Robinson, J. S. Burgess, C. E. Junkermeier, S. C. Badescu, T. L. Reinecke, F. K. Perkins, Maxim K. Zalalutdniov, J. W. Baldwin, J. C. Culbertson, P. E. Sheehan, and E. S. Snow, Nano Lett. {\bf 10}, 3001 (2010).

\bibitem{Shenoy}A. Bagri, C. Mattevi, M. Acik, Y. J. Chabal, M. Chhowalla, and V. B. Shenoy, Nat. Chem. {\bf 2}, 581 (2010).

\bibitem{Balakrishnan}J. Balakrishnan, G. K. W. Koon, M. Jaiswal, A. H. Castro Neto, and B. \"Ozyilmaz, Nat. Phys. {\bf 9}, 284 (2013).

\bibitem{Wagner}P. Wagner, V. V. Ivanovskaya, M. Melle-Franco, B. Humbert, J.-J. Adjizian, P. R. Briddon, and C. P. Ewels, Phys. Rev. B {\bf 88}, 094106 (2013).

\bibitem{Ramasubramaniam}A. Ramasubramaniam, D. Naveh, and E. Towe, Nano Lett. \textbf{11}, 1070 (2011).

\bibitem{JACS}P. Lazar, F. Karlick\'y, P. Jurecka, K. Kocman, E. Otyepkov\'a, K. Saf\'aro\'a, M. Otyepka, J. Am. Chem. Soc. {\bf 135}, 6372 (2013).

\bibitem{Denis}P. A. Denis, Phys. Chem. Chem. Phys., DOI: 10.1039/C4CP05331A.

\bibitem{ezawa}M. Ezawa, Phys. Rev. Lett. {\bf 109}, 055502 (2012).

\bibitem{n6}C.-C. Liu, W. Feng, and Y. Yao, Phys. Rev. Lett. \textbf{107}, 076802 (2011).

\bibitem{n7}L. Matthes and F. Bechstedt, Phys. Rev. B \textbf{90}, 165431 (2014).

\bibitem{kaloni-prb}T. P. Kaloni, S. Gangopadhyay, N. Singh, B. Jones, and U. Schwingenschl\"ogl, Phys. Rev. B {\bf 88}, 235418 (2013).

\bibitem{Ulbricht}R. Zacharia, H. Ulbricht, and T. Hertel, Phys. Rev. B \textbf{69}, 155406 (2004).

\bibitem{PRB-germanene}Y. Cai, C.-P. Chuu, C. M. Wei, and M. Y. Chou, Phys. Rev. B \textbf{88}, 245408 (2013).

\bibitem{ass} M. Houssa, B. van den Broek, E. Scalise, B. Ealet, G. Pourtois, D. Chiappe, E. Cinquanta, C. Grazianetti, M. Fanciulli, A. Molle, V. V. Afanasev, and A. Stesmans, Appl. Sur. Sci. \textbf{291}, 98 (2014).

\bibitem{Li}L. Li, S.-Z. Lu, J. Pan, Z. Qin, Y.-Q. Wang, Y. Wang, G.-Y. Cao, S. Du, and H.-J. Gao, Adv. Mater. \textbf{26}, 4820 (2014).

\bibitem{Bampoulis}P. Bampoulis, L. Zhang, A. Safaei, R. V. Gastel, B. Poelsema, and H. J. W. Zandvliet, J. Phys.: Condens. Matter \textbf{26}, 442001 (2014).

\bibitem{Hapala}M. $\check{\rm S}$vec, P. Hapala, M. Ondr\'a$\check{\rm c}$ek, P. Merino, M. Blanco-Rey, P. Mutombo, M. Vondr\'a$\check{\rm c}$ek, Y. Polyak, V. Ch\'ab, J. A. Mart\'in Gago, and P. Jel\'inek, Phys. Rev. B \textbf{89}, 201412 (2014).

\bibitem{physicae}M. Ye, R. Quhe, J. Zheng, Z. Ni, Y. Wang, Y. Yuan, G. Tse, J. Shi, Z. Gao, and J. Lu, Physica E \textbf{59}, 60 (2014).

\bibitem{pccp-14}S.-S. Li, C.-W. Zhang, W.-X. Ji, F. Li, P.-J. Wang, S.-J. Hu, S.-S. Yan, and Y.-S. Liu, Phys. Chem. Chem. Phys. \textbf{16}, 15968 (2014).

\end{thebibliography}
\end{document}